# Perspectives on risk prioritization of data center vulnerabilities using rank aggregation and multi-objective optimization

Bruno Iochins Grisci[a,b,1], Gabriela Kuhn[a,1], Felipe Colombelli[a,b,1], Vítor Kehl Matter[a,1], Leomar Lima[c], Karine Heinen[c], Mauricio Pegoraro[c], Marcio Borges[c], Sandro José Rigo[a], Jorge Luis Victória Barbosa[a], Rodrigo da Rosa Righi[a], Cristiano André da Costa[a] and Gabriel de Oliveira Ramos[a,*]

[a]*Software Innovation Laboratory - Softwarelab, Graduate Program in Applied Computing, Universidade do Vale do Rio dos Sinos, Av. Unisinos, 950, São Leopoldo, 93022-750, RS, Brazil*
[b]*Institute of Informatics, Federal University of Rio Grande do Sul, Av. Bento Gonçalves, 9500, Porto Alegre, 91501-970, RS, Brazil*
[c]*Dell Technologies Brazil, Av. Industrial Belgraf, 400, Eldorado do Sul, 92990-000, RS, Brazil*

## ARTICLE INFO

*Keywords*:
vulnerabilities
multi-objective optimization
database security
risk prioritization
rank aggregation

## ABSTRACT

Nowadays, data has become an invaluable asset to entities and companies, and keeping it secure represents a major challenge. Data centers are responsible for storing data provided by software applications. Nevertheless, the number of vulnerabilities has been increasing every day. Managing such vulnerabilities is essential for building a reliable and secure network environment. Releasing patches to fix security flaws in software is a common practice to handle these vulnerabilities. However, prioritization becomes crucial for organizations with an increasing number of vulnerabilities since time and resources to fix them are usually limited. This review intends to present a survey of vulnerability ranking techniques and promote a discussion on how multi-objective optimization could benefit the management of vulnerabilities risk prioritization. The state-of-the-art approaches for risk prioritization were reviewed, intending to develop an effective model for ranking vulnerabilities in data centers. The main contribution of this work is to point out multi-objective optimization as a not commonly explored but promising strategy to prioritize vulnerabilities, enabling better time management and increasing security.

## Contents

*Corresponding author
gdoramos@unisinos.br (G.d.O. Ramos)
ORCID(s): 0000-0003-4083-5881 (B.I. Grisci); 0000-0003-1718-3109 (G. Kuhn); 0000-0002-8684-8955 (F. Colombelli); 0000-0001-9422-0757 (V.K. Matter); 0000-0001-8140-5621 (S.J. Rigo); 0000-0002-0358-2056 (J.L.V. Barbosa); 0000-0001-5080-7660 (R.d.R. Righi); 0000-0003-3859-6199 (C.A.d. Costa); 0000-0002-6488-7654 (G.d.O. Ramos)
[1]These authors contributed equally to this work.

## 1. Introduction

Vulnerabilities are flaws in systems, processes, and strategies that result in risks [1]. Regarding vulnerabilities in software, Le et al. [2] defined them as security bugs that impair the confidentiality, integrity, or availability of software systems. Considering the information security scenario, data centers often face situations where it is impossible to fix all the detected vulnerabilities. In cases where resources or time are scarce, having the vulnerabilities ranked by risk would be essential to validate which vulnerabilities should be patched first. In particular, it becomes fundamental to rank the vulnerabilities and servers to receive patches effectively and protect the data and infrastructure, ensuring that the most critical issues are handled quickly. Further research can significantly improve this growing area [2, 3], and it is imperative to discover optimized strategies to prioritize vulnerabilities.

According to a 2019 report survey [4], unpatched vulnerabilities remain the leading cause of today's most serious data breaches. Data collected from 340 infosecurity professionals indicated that 39% of the companies scan vulnerabilities monthly or less often than that. A significant group (27%) reported that they did not fix vulnerabilities in a month or less, and approximately half reported not applying patches in two weeks or less. Another report from 2020 [5] put the average cost of a data breach at $3.86 million and indicated that proper incident response preparedness resulted in a $2 million reduction in costs.

The list of known information security vulnerabilities increases every day. The Common Vulnerabilities and Exposures (CVE) system, which provides a reference method for publicly known vulnerabilities and exposures, reported 18,325 new vulnerabilities in 2020 and 20,141 in 2021[2]. When a new vulnerability is identified, the CVE database or local databases are updated depending on the case. Building upon such databases, scanning tools can detect the vulnerabilities in servers when they exist. Once the scanning step finishes, all discovered vulnerabilities are supposed to be handled by the security team. Nevertheless, if the amount of vulnerabilities surpasses the team's human, time, or financial resources, a risk prioritization strategy is needed for dealing with the highest risk vulnerabilities. The Common Vulnerability Scoring System (CVSS) is a standard that classifies the vulnerabilities' security severity. However, as indicated by Holm et al. [6], modeling security factors only with CVSS data does not accurately represent the time-to-compromise of a system, creating the need for systems that consider all vulnerabilities.

In this context, the main goal of this article is to describe how to make intelligent decisions regarding which server and vulnerabilities should receive attention first. Relying on a single metric to rank the vulnerabilities may fail to account for the underlying complexity of the problem. Moreover, considering a dynamic environment, even the use of multiple metrics to create a global metric that tries to capture the richness of the information is insufficient, as a single model may fail to consider changes of context or shifts in the users' needs [7]. A promising approach to solve this problem, thus, is the use of multi-objective optimization to obtain a set of optimal vulnerabilities ranks that consider multiple metrics simultaneously. The main contributions of this article are an updated survey about vulnerabilities risk prioritization and an entry point for ranking methodologies and multi-objective optimization applied to this domain. All discussions present in this work are application and technology agnostic and can be applied to different vulnerabilities prioritization scenarios. This paper reviews the literature; therefore, no experiments were performed. Perspectives and new research directions are backed up by theoretical and practical arguments discussed in recent publications.

To the best of our knowledge, this is the first work to review published vulnerabilities risk prioritization research through the theoretical *ranking* framework and to propose the use of multi-objective optimization. Le et al. [2] presented a taxonomy of software vulnerability studies and discussed the present limitations and potential solutions to manage vulnerabilities, though their survey neither specified ranking nor multi-objective methodologies. Lin [8], Deng et al. [9], and Li et al. [10] reviewed rank aggregation methods, and Roijers and Whiteson [11], Hayes et al. [12], and Saini and Saha [13] reviewed multi-objective methods, but they neither focused on risk prioritization nor security.

Multi-objective optimization is not often used for vulnerability risk prioritization, but notable exceptions exist. The framework by Farris et al. [14] used multi-objective to select the vulnerabilities to be patched among the ones present in the considered infrastructure, while Viduto et al. [15] used the multi-objective optimization to investigate risk assessment cost-effectively. Jacobs et al. [16] and Beck and Rass [17] recognized that risk prioritization should not be handled using a single metric or objective but do not discuss the perspective of a multi-objective approach.

The remainder of this article presents the survey results in Sections 2 to 4 and a proposition in Section 5. Section 2 introduces several topics relevant to risk prioritization, such as the most common metrics and approaches for risk estimation and commonly used datasets. Section 3 presents the problem of risk prioritization from the perspective of ranks and discusses related issues. Section 4 introduces multi-objective optimization and how it can be applied to risk prioritization. Finally, in Section 5, the research perspectives are discussed.

## 2. Vulnerability risk assessment

Computing the security risk in an environment represents essential information to companies trying to mitigate server attacks. Identifying risk scenarios and providing feedback of the server vulnerabilities' urgency can assist security teams to prioritize the remediation efforts based on their organizational structure's needs and goals. This section discusses academic and commercial efforts towards these goals. Figure 1 illustrates a general vulnerability management life cycle that considers the infrastructure, vulnerabilities detection, prioritization strategies, and vulnerabilities management. A complete description of the whole vulnerability management practice can be found in the work by Foreman [1].

The advantages of the vulnerability management life cycle are that it contemplates the necessity of a scoring system (one of the most relevant metrics for scoring vulnerabilities is described in Section 2.1) and that it enables the evaluation of vulnerabilities risk (e.g., one of the scoring algorithms detailed in Section 2.2 can be employed to accomplish it). The attributes used to score vulnerabilities are discussed in Section 2.3. This proposed scheme is considered efficient because it enables vulnerabilities prioritization. This section presents several ways by which vulnerabilities can be prioritized and have their risk assessed. In contrast, Sections 3 and 4 show how different results can be combined into a more robust and flexible recommendation.

---

[2] https://www.cvedetails.com/browse-by-date.php

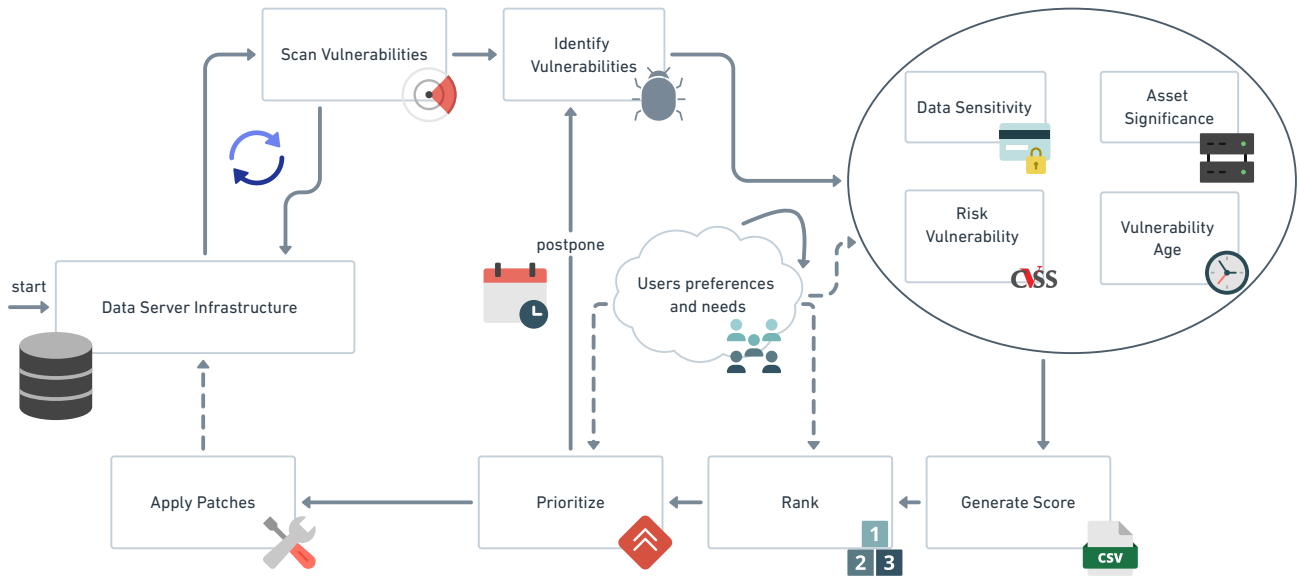

**Figure 1: Vulnerability management life cycle.** The cycle starts with a scan application that monitors the data server infrastructure of the organization and detects vulnerabilities. These vulnerabilities are then classified according to distinct metrics, such as their age and risk, as well as the relevant security attributes of the assets in which they were identified, such as the data sensitivity. A general score is given to each vulnerability from these metrics, which enables their ranking and prioritization. For instance, the ranking step can employ rank aggregation or multi-objective methodologies. Due to time and resources constraints, not all vulnerabilities can be patched immediately, so the prioritized ones are fixed, and the remaining ones are postponed. The vulnerability patching may, on occasion, accidentally introduce new vulnerabilities in the infrastructure. At any given time, the users can redefine the relevant vulnerabilities attributes and assets that should be considered, create ranks, and prioritize each vulnerability.

### 2.1. Common Vulnerability Scoring System

The CVSS[3] is an open framework for providing the properties and severity of software vulnerabilities. This score is a common approach for prioritizing vulnerability remediation, furnishing a numerical score reflecting the vulnerability severity, which can be combined with other factors to help organizations prioritize their vulnerabilities appropriately.

The CVSS consists of three metric groups: Base, Temporal, and Environmental. The *Base metric* represents the vulnerability's intrinsic characteristics constant over time and across user environments. The *Temporal metric* reflects the characteristics of a vulnerability that may change over time but not across user environments. Finally, the *Environmental metric* represents the characteristics of a vulnerability that are relevant and unique to a particular user's environment.

The CVSS Base metric produces a score between 0.0 and 10.0, in which 10.0 is the most severe level. The Temporal and Environmental metrics can then adjust the Base score based on specialized knowledge such as the time-varying properties, availability of exploit code, and the presence of mitigations in that environment. Scoring the Temporal and Environmental metrics is not required but is recommended for more precise scores[4].

Historical vulnerability information is of great importance for vulnerability assessment. However, collecting information about all known security flaws is not simple. Although thousands of sources generate vulnerability information [18], such data is often not publicly and freely available due to the security implications. Because of these challenges, public databases were organized to provide identifiers and other information for known security weaknesses. One such initiative that stands out is the Common Vulnerabilities and Exposures (CVE)[5] project provided by the MITRE Corporation[6] and the National Vulnerability Database (NVD)[7] repository maintained by the National Institute of Standards and Technology (NIST)[8], U.S. Department of Commerce. Beyond providing identifiers, descriptions, and other references, NVD also provides CVSS scores for the vulnerabilities.

---

[3] https://www.first.org/cvss/
[4] https://www.first.org/cvss/specification-document
[5] https://cve.mitre.org
[6] https://mitre.org
[7] https://nvd.nist.gov
[8] https://nist.gov/

**Table 1**
Summary of works proposing methods for computing vulnerabilities risk. The works that do not contain a specific approach, denoted by −, possess their own set of strategies.

| Reference | Proposed Method | Approach | Target | Goal |
| --- | --- | --- | --- | --- |
| Beck and Rass [17], 2016 | Automated CVSS Risk Aggregation | Neural Networks | Enterprise Infrastructure | To propose a method that resembles human experts' decision making in risk assessments. |
| Farris et al. [14], 2018 | VULCON | Goal Programming | Cyber-security Operations Center | To offer guidance to improve vulnerability response processes. |
| Jacobs et al. [16], 2019 | EPSS | Logistic Regression | Information System | To analyze the probability of a vulnerability being exploited in the wild within the first twelve months after public disclosure. |
| Jiang et al. [23], 2012 | VRank | − | Service Oriented Architecture | To score and rank vulnerabilities considering intrinsic properties and services contexts that contain the vulnerability. |
| Singh and Joshi [24], 2016 | Hazard Metric | − | Network Environment | To identify the probability of attacks in user environments. |
| Spanos et al. [25], 2013 | WIVSS | − | Information System | To score vulnerabilities, depending on the different impact of vulnerabilities characteristics. |
| Viduto et al. [15], 2012 | RAOM | Multi-objective Tabu Search | Network Environment | To solve a security countermeasure selection problem, assessing risks and cost-effectively minimizing them. |
| Wang et al. [26], 2010 | OVM | Ontology | Software Products | To provide a set of security metrics to rank attacks based on vulnerability analysis. |
| Yin et al. [19], 2020 | ExBERT | BERT | Software | To predict if a vulnerability will be exploited or not. |
| Zeng et al. [27], 2021 | LICALITY | Neural Networks | Network Environment | To capture the attacker's preference on exploiting vulnerabilities. |

Although the industry uses the CVSS as a *de facto* standard to measure cybersecurity risk [19], CVSS scores have some weaknesses in assessing the overall severity of security vulnerabilities. Predicting the exploitability of vulnerabilities indicates the level of a potential attack, and CVSS has limitations for calculating such risk [20]. Moreover, the time delay between the publication of vulnerabilities and the availability of their CVSS scores is a historical problem [21]. Also, the highly specialized knowledge necessary to calculate CVSS scores [22] is a concern.

Furthermore, the vulnerability prioritization within an organization network can also be linked to other factors beyond the vulnerability characteristics. Recognizing security demands and considering them in conjunction with CVSS is necessary to prioritize them according to each company's purpose.

## 2.2. Computing vulnerabilities risk

In addition to the CVSS score, other factors can be considered for computing the real threat of an attack. The vulnerability age, the availability of exploit code, malware kits, and the server environment are examples of attributes to be included in the risk prioritization, giving the organizations a guide to focus the remediation efforts based on real-risk exploitability. A summary of the works related to vulnerability risk is shown in Table 1, listed by the authors' names in alphabetical order and described in more detail in this section. The proposed methods, approaches, targets, and goals brief descriptions are presented to assist the understanding of what differentiates them.

Avoiding wasting time patching vulnerabilities that could be delayed is the solution proposed by Jacobs et al. [16] using the framework Exploit Prediction Scoring System (EPSS) to prioritize the vulnerabilities. The EPSS assesses vulnerability threats by analyzing the probability of a vulnerability being exploited within 12 months of being publicly disclosed. The main goal is to ensure that most exposed vulnerabilities will be fixed first. The framework validates a standard logistic regression considering a feature selection process that computes the Precision/Recall Area Under the Curve (PR AUC) and the Bayesian Information Criteria (BIC).

Based on the six factors used by the CVSS Base metric (Access Vector, Access Complexity, Authentication, Confidentiality Impact, Integrity Impact, and Availability Impact) and suggesting that Integrity, Availability, and Confidentiality Impact have different levels of importance in security, Spanos et al. [25] introduced a new vulnerability scoring system called Weighted Impact Vulnerability Scoring System (WIVSS), which considers different weights to the impact metrics compared to the CVSS, reflecting this distinction. The WIVSS score is derived from mathematical approximations that use the CVSS Base metrics considering a rounded decimal value obtained through maintaining the three metrics that define the exploitability with the same weights as CVSS, only changing the impact metrics weights.

Many factors such as the vulnerabilities' maturity, frequency, and impact on the system increase the risk score related to network architecture and to how a vulnerability behaves on the system affects the organization's security risk level [24]. Singh and Joshi [24] proposed a method to quantify the security risk considering such factors. The proposed Hazard Metric uses CVSS as the base score and considers specific user information to aggregate metrics to measure the probability of exploiting the vulnerability using the critical resources of an organization network.

The Hazard Metric estimates the security strength of a specific network based on the following five metrics. The Maturity Level defines how long the vulnerability has been present in the network system. The Frequency of Exploit computes the likelihood of the vulnerability being exploited in the user's environment. The Exploitability Impact defines the impact that the exploitation of a vulnerability would have on a network configuration. The Amendment Level measures the degree of resistance against a vulnerability. And the Authentication Level determines the level of privileges required by an attacker.

Another example elucidating the importance of considering the environment is the Service-Oriented Architecture (SOA) context [23]. Several services in SOA are built on diverse hardware and software platforms and offered by different providers. It makes the vulnerabilities faced by SOA much more diverse than those faced by traditional software. Also, the impact of a specific vulnerability can be different in the SOA process than the suggested by the CVSS score. Taking that into account, Jiang et al. [23] designed the VRank framework for scoring and ranking vulnerabilities for SOA services. The VRank framework uses a dependency graph to capture the dependency relation between service components concerning a security requirement and estimates a service's importance in a business process. Afterward, two scoring functions that use the vulnerabilities exploitability significance, threat levels, and other properties are employed to acquire the vulnerability score.

The process of how risk is aggregated is often context-dependent and rarely documented and is not appropriately studied from a psychological perspective [17]. In many contexts, the work is done using rules-of-thumb. Such an approach wastes information, and often security goals are managed manually. Consequently, the security experts need to refine the risk aggregation using their expertise. Aiming to support vulnerabilities prioritization and decision, Beck and Rass [17] proposed an approach using neural networks to deal with a particular security circumstance resembling a human expert's decision-making in the same regard. An evaluation of 13 neural networks was proposed using an automated aggregation. It would acquire the risk aggregation through hierarchical aggregation to feed a CVSS risk assessment with meta-metric into the neural networks and then reuse its output as input to the neural networks in the next stage.

The "LIkelihood and critiCALITY" (LICALITY) risk prioritization system also uses neural networks, in addition to probabilistic logic programming [27]. LICALITY models the threat scenarios in network environments and tries to capture attackers' preferences on how to exploit vulnerabilities. The vulnerability's risk is defined by the criticality of exploitation and the likelihood of exploitation. The authors reported a reduction of vulnerability remediation work compared to the CVSS by a factor between 1.85 and 2.89.

Focusing on mitigating higher risk attacks on time, Wang et al. [26] studied a new approach for computing the risk attacks against systems. The Ontology for Vulnerability Management (OVM) was proposed as a knowledge base and data source for calculating the attacks' rank using relationships among CVE, CWE[9], CVSS, and CAPEC[10]. The ontology includes concepts such as Vulnerability, Information Technology (IT) Product, Attacker, Attack, Consequence, and Countermeasure, which retrieve all the vulnerability information concerning a software product. Therefore, it also creates the attack patterns by classifying all vulnerabilities based on 14 predefined types to rank the vulnerabilities.

Wang et al. [26] also considered that the newest vulnerabilities require more attention as fewer patches exist for them compared to old vulnerabilities. Thus, the work divided the vulnerability disclosure history into three time intervals: *present*, *recent*, and *past*. For each time interval, a weight that reflects their importance was assigned. Therefore, the *present* interval was considered more important than the *recent* interval, while the *recent* is more significant than the *past* interval. The algorithm calculates a weight for each type of vulnerability based on: (1) the time of discovery, whether it was discovered long ago or recently; (2) frequency, which measures the number of vulnerabilities of this same type; and (3) severity, given by the CVSS vulnerability scores. The final attack pattern weight is the sum of related vulnerability types' weights, which ranks the attacks.

Every year, the number of vulnerabilities disclosed to the public has been increasing the efforts and resources required to collect and maintain the vulnerabilities datasets. Since the chance of a vulnerability being exploited increases by up to 5 orders of magnitude once it is disclosed [28], efficient data management and extraction are critical to prioritizing efforts. Based on vulnerabilities description contents, which are easy to find and rich in semantic information, Yin et al. [19] proposed to train a model called ExBERT to predict the exploitability of vulnerabilities using transfer learning to help the experts prioritize the patch application. ExBERT mainly consists of two stages: Bidirectional Encoder Representations from Transformers (BERT) transfer learning and exploitability prediction application. BERT transfer learning means fine-tuning a pre-trained BERT [29] into a fine-tuned BERT model using a collected description corpus. Exploitability prediction application deals with wordpiece tokenization, token embedding, sentence embedding, and exploitability prediction.

By investigating risk assessment methodologies in networks to improve decision-making approaches, Viduto et al. [15] provided a tool to select security parameters taking more than one objective into account. By applying a multi-objective algorithm (Multi-objective Tabu Search, described in Section 4.1), the Risk Assessment and Optimization Model (RAOM) consists of two processes: risk assessment and an optimization routine. Therefore, the model was built to provide a method for assessing risks and cost-effectively minimizing them through security countermeasures. The RAOM divides the risk assessment process into eleven steps: (1) Identify organizations' essential functions; (2) Identify essential systems; (3) Assess systems for vulnerabilities; (4) Analyse vulnerabilities; (5) Analyse vulnerability properties; (6) Verify the vulnerability to attacks; (7) Impact analysis; (8) Threat-vulnerability analysis; (9) Likelihood determination; (10) Risk level determination; and (11) Security control recommendation. The second process introduces the Multi-objective Tabu Search, which is used to find optimal solutions cost-effectively.

---

[9] Common Weakness Enumeration - https://cwe.mitre.org
[10] Common Attack Pattern Enumeration and Classification - http://capec.mitre.org/

Another strategy is investigated by Farris et al. [14]. The authors propose two metrics called Total vulnerability exposure (TVE) and Time-to-vulnerability remediation (TVR). The TVE is computed based on a mitigation utility function, defined as the weighted sum of the vulnerability CVSS severity, the number of months a vulnerability has persisted without mitigation, and the underlying vulnerability's chronological age. The weights of these attributes are user-defined and given as input to their VULCON framework. The TVE is calculated after an optimization process (discussed in more detail in Section 4.1) by computing the sum of the mitigation utility values of the vulnerabilities not selected in the optimization step. On the other hand, the TVR metric is used to access how persistent a vulnerability is in the scanned system. It is calculated as the difference between the vulnerability's detection time and the mitigation selection time. The TVR is important since it is impossible to mitigate all vulnerabilities present in an extensive infrastructure periodically scanned [14]. The authors present this metric as the tolerance indicator for the latency between discovery and response.

Because of the value data centers represent for companies, commercial software also focuses on evaluating vulnerability risk. These tools focus on providing IT and security teams with an understanding of the vulnerabilities' urgency to prioritize the remediation efforts. Nexpose[11], Kenna Security[12], Intruder[13], and Nessus[14] are examples of risk-based vulnerability management systems. They generate a score on which the vulnerabilities are classified per risk level. Nonetheless, their internal methods and metrics are private, preventing a deeper scientific analysis. Moreover, none of these tools enable a true multi-objective prioritization of vulnerabilities.

As seen in this section, there are several different ways to approach the risk assessment problem. Each reviewed work presented significant contributions. However, the potential of multi-objective methodologies for vulnerability prioritization is not completely exploited, so more challenges will be discussed in Section 5.

### 2.3. Attributes for risk prioritization

As mentioned in the previous sections, several attributes could be used as metrics or objectives to determine the risk level of a given vulnerability. Some of them are represented as examples in Figure 1. The CVSS severity described in Section 2.1 and its variations are the main attributes used for risk prioritization, while each work from Section 2.2 proposes valid alternatives.

There is no single set of attributes capable of representing risk as a whole [16], and ideally, several of them should be considered simultaneously according to the current user needs. Some attributes suggested by Farris et al. [14] included vulnerability persistence (for how long it remains present in the assets without being patched), vulnerability age (for how long it has been known to the public), and how many hours are needed to apply a patch. Singh and Joshi [24] suggested using the likelihood of an exploit happening, the exploit impact, the system resistance against a vulnerability, and the privilege level needed by an attacker to successfully exploit the vulnerability. Attributes can also represent the risk of the organization assets (rather than of vulnerabilities themselves) in which vulnerabilities are present. For instance, Jiang et al. [23] proposed the importance of the asset or of its application as relevant information to measure risk. Similarly, the importance or confidentiality of the data stored in servers where vulnerabilities were detected can be a reference for risk estimation. Sections 3 and 4 introduce different ways in which these attributes can be used simultaneously.

## 3. Ranking for risk priorities

The problem of risk prioritization can be represented as a rank. A *rank* is an ordered list of items or elements that indicates a preference order. A *ranking* is a position of an item in a rank, represented by a number, such that the first item has the lowest ranking (usually valued 1). Usually, the preferable item has the first (the lowest) ranking in a rank. For instance, taking vulnerabilities as items in a list and ordering them from the most severe to the less severe risk would result in a rank of vulnerabilities prioritizing risk. In this example, the vulnerability with the highest risk would be the first item of the rank and receive the lowest ranking (for instance, 1).

Ranks can be classified concerning their data structure, represented by pairs of keys and values (Figure 2a). If it is item-based, each key represents an item, and each value represents the item's ranking. If it is rank-based, the keys represent the ranking, and the values are the ranking's items. If multiple ranks of the same items are built from different metrics or rankers, it is useful to consider them together in a matrix. Each row value is an item (rank-based) or a ranking (item-based) in this matrix, and each column is the rank corresponding to a specific metric. The item-based format has some advantages over the rank-based, especially when items are tied (two or more items have the same ranking). With item-based, tied items can receive the same ranking value, but in the case of rank-based, the items must be arbitrarily broken because all keys must be unique [10]. This is particularly undesirable for risk prioritization, a case in which several ties are expected because the same vulnerabilities can appear in different assets.

---

[11]https://www.rapid7.com/products/nexpose/
[12]https://www.kennasecurity.com/
[13]https://www.intruder.io/
[14]https://www.tenable.com/products/nessus

There are different ways to represent ties in an item-based rank, four of which are shown in Figure 2b and described below.

- `min`: each tied item receives the minimal ranking available, and the next ranking considers the number of tied items;

- `max`: each tied item receives the maximum ranking available, and the next ranking considers the number of tied items;

- `dense`: each tied item receives the minimal ranking available, and the next ranking ignores the number of tied items;

- `mean`: each tied item receives the available rankings mean, and the next ranking considers the number of tied items.

The `mean` representation has two advantages. It preserves the positioning of the untied items, so in the example in Figure 2b in which there are two items tied for second place, this information is not lost in the ranking of the next item, which will be 4, signaling the absolute number of items ranked better than it. It also "punishes" tied items by giving them an intermediate ranking value, so it is possible to know if an element is tied or not when comparing two ranks. In the previous example, we know that there is a tie in the second position because the ranking of the items would be 2.5, while in a rank without ties, the second item would have a ranking of 2.

For instance, a rank of vulnerabilities could be created by ordering the vulnerabilities by their CVSS score. However, using only one metric or *measure of security* is insufficient to capture the complexity, high-dimensionality, and non-linearity of risk assessment [17]. One solution, thus, is to use multiple metrics that capture different aspects of what defines *risk*. In this case, if there are $n$ metrics, ordering the vulnerabilities by each metric will create $n$ distinct ranks. This becomes the problem of combining several ranks to obtain a single risk rank, which is more reliable than the base rankers [10]. The following section will discuss rank aggregation, the traditional approach to combine multiple ranks, and Section 4 will focus on the multi-objective approach for the problem.

### 3.1. Rank aggregation

Suppose there are multiple ranks of the same items available. For instance, if each rank represents vulnerabilities ordered by a distinct metric of risk factor, as discussed in

**Table 2**
**Example of ranks and the Arrow theorem.** The table below shows three ranks (1, 2, and 3), each of them ranking three items (a, b, and c).

| Individual rank | First item | Second item | Third item |
|---|---|---|---|
| 1 | *a* | *b* | *c* |
| 2 | *b* | *c* | *a* |
| 3 | *c* | *a* | *b* |

Section 2.3, it may be desirable to combine all these individual ranks into a single global rank to obtain a multifaceted representation of the risk. Combining different ranks of the same elements is known as rank aggregation [8, 10]. One of its first uses was in the social and political sciences as an election strategy. In this case, each individual would rank their vote preferences, and the election winner would be determined by aggregating all these lists. In recent years, rank aggregation methods have shown to be helpful in several fields, especially for analyzing high-throughput, omics-scale, biological data [8, 10]. Any algorithm combining different ranks to produce a global rank should usually satisfy the following three properties:

- **Non-dictatorship:** the algorithm cannot always select one of the individual's ranks.

- **Unanimity:** if every individual ranker prefers item *a* to item *b*, the global rank must prefer *a* to *b*.

- **Independent of irrelevant alternatives:** if individual ranks are modified, but the order of items *a* and *b* is unchanged, the global order of *a* and *b* should not change.

However, the result of Arrow's impossibility theorem [30] shows that any algorithm capable of creating a global rank of at least three elements that satisfy the *unanimity* and *independent of irrelevant alternatives* properties is necessarily a dictatorship. In other words, it is theoretically impossible to satisfy all three properties simultaneously.

Table 2 shows an example in which aggregating three distinct ranks of three items fails. If an algorithm ranks the items by comparing *a* to *b* and then comparing *b* to *c*, when comparing *a* to *b*, two of the individual ranks prefer *a* to *b*. And, when comparing *b* to *c*, two individual ranks also prefer *b* to *c*. By transitivity, it would be assumed that the individual ranks would prefer *a* to *c*, but it is not the case. The conclusion is that *a* is preferred to *b*, *b* is preferred to *c* and *c* is preferred to *a*.

Nevertheless, rank aggregation is still beneficial, as mentioned before, and the development of algorithms for aggregation is an active and rich field of study [8]. As defined by Li et al. [10], rank aggregation methods can be divided into the following categories:

- **Bayesian methods**: use quantities involved in posterior inference, such as posterior probability or Bayes factor, to determine the aggregated rank. Examples:

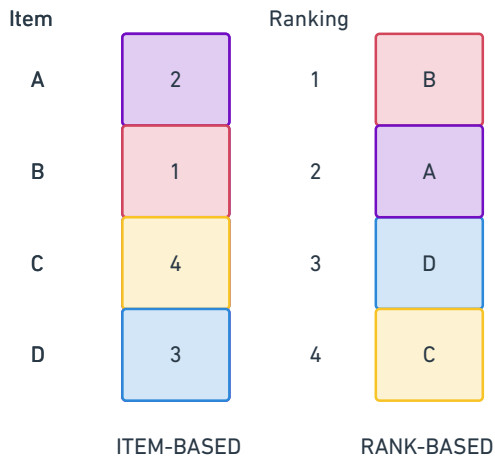
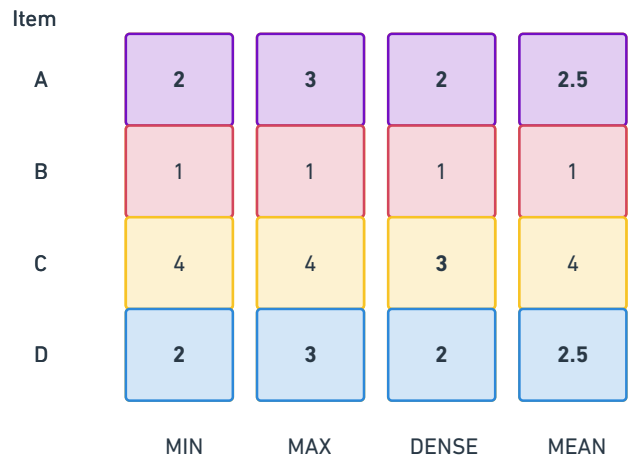

**Figure 2: Distinct rank representations.** a) Two arrangements of the same rank with four items (A, B, C, and D, each with a unique color). In the item-based representation, the list contains the ranking of each item, so that A is the second item in the rank, B is the first item, and so forth. For the rank-based, the order in which the items appear in the list is the rank order. Note that it is impossible to represent ties in this case because each ranking corresponds to a single index of the list. b) Four distinct ways to represent ties with the item-based rank. In this example, items A and D are tied in the same position behind item B and in front of item C. The rankings changing in each representation are marked in bold.

Bayesian Aggregation of Rank Data (BARD) [9], Bayesian Iterative Robust Rank Aggregation (BIRRA) [31], Bayesian Aggregation of Rank-data with Covariates (BARC) [32].

- **Non-optimization-based methods**: use summary statistics such as the arithmetic mean, median, geometric mean, and L2-norm to aggregate ranks. Examples: Borda's collection [8].

- **Optimization-based methods**: minimize a distance measure (Section 3.2) to obtain a final rank as close as possible to all individual ranks. Examples: Cross Entropy Monte Carlo (CEMC) methods [33].

- **Distribution-based methods**: use a probabilistic latent model or distributional information of any statistic calculated from the rank data. Examples: Thurstone's model [34], Robust Rank Aggregation (RRA) [35].

- **Markov chain-based methods**: use a Markov Chain (MC) modeling framework, where the union of items from the individual ranks creates the state space. Examples: MC1, MC2, MC3 [8, 36].

An in-depth description of the above categories and comparisons between them can be found in the works by Lin [8], Deng et al. [9], and Li et al. [10]. For risk prioritization, the problem is considered best represented by the globally full list scenario [10], in which every item is explicitly ranked by all the rankers (distinct metrics). This is the ideal case for rank aggregation algorithms [10]. Risk prioritization would usually fall under the "a few long ranked lists" representation, as the ranks can have thousands of vulnerabilities but a relatively small number of risk metrics to consider. This is important because Markov chain methods, CEMC methods, RRA, Stuart, BARD, and BIRRA are expected to perform better for "a few long lists", while Thurstone's models, paired comparison models, and multistage models require "many short lists" [8, 36]. Borda's methods have been shown to work reasonably well for both data structures [10]. However, this is not often well documented nor comprehensively studied regarding the problem of risk aggregation [17]. This leads the process of aggregation to be imprecisely defined and to rely heavily on rules-of-thumb [17]. The process is also not unified, so the adopted practices often change for each application domain and depend on context.

### 3.2. Distance metrics

A significant requirement when dealing with ranks is how to compare them. The distance or similarity between candidate ranks or between a proposed aggregated rank and the individual ranks can be used to build objective functions in optimization algorithms. For this purpose, the literature describes distance metrics that measure how different two ranked lists are from each other. This section describes three of the most used metrics, the Canberra distance, the Spearman's rank correlation coefficient, and the Kendall tau rank distance.

The Canberra distance is defined as the sum of differences between the two ranked lists divided by the sum of their position in the rank [37]. Its values start at zero (the ranked lists are equal) and increase for every non-matching pair of elements. For two given ranks $p$ and $q$ encoded as permutation lists, the Canberra distance is defined as:

$$D_{canberra}(p, q) = \sum_{i=1}^{n} \frac{|p_i - q_i|}{|p_i| + |q_i|}, \quad (1)$$

where $p_i$ is the position of the $i$-th item in rank $p$, and $q_i$ is the position of the $i$-th item in rank $q$. One of its main advantages is that it gives higher values for mismatches at the top of the rank that are usually more significant. When $p_i$ and $q_i$ are small, that means that the item in consideration is highly ranked. Therefore, the ranking difference is divided by their sum, yielding higher values for more essential items than those ranked at the bottom of the rank.

The Spearman's rank correlation coefficient (also called Spearman's $\rho$) measures the correlation between two random variables. The Spearman's $\rho$ between two ranked lists is equal to Pearson's correlation [38] between the rank values of the two ranks, so the two metrics have the same boundaries. It results in values between $-1$ and 1, with 1 indicating that both variables are linearly correlated and $-1$ meaning that they are linearly inversely correlated. A zero result means that there is no correlation between the ranks. In the case where there are no ties in the ranking, Kalousis et al. [39] defined the formula of Spearman's $\rho$ as in Equation 2, where $p$ and $q$ are two ranks presented as permutation lists, and $n$ is the number of elements in the ranks.

$$r_s(p, q) = 1 - \frac{6 \sum_{i=1}^{n}(p_i - q_i)^2}{n(n^2 - 1)} \quad (2)$$

The Kendall tau rank distance [40] measures the correspondence between two ranks by counting the number of pairwise disagreements between them. It can be defined by Equations 3 and 4, in which $\tau_1(i)$ and $\tau_2(i)$ are the rankings of element $i$ in the lists $\tau_1$ and $\tau_2$, and $P$ is the set of unsorted pairs of distinct elements in $\tau_1$ and $\tau_2$. The normalized Kendall tau rank distance returns a value between 1 (identical lists) and $-1$ (one of the lists is the reverse of the other).

$$K(\tau_1, \tau_2) = \sum_{\{i,j\} \in P} \bar{K}_{i,j}(\tau_1, \tau_2) \quad \bar{K}_{i,j}(\tau_1, \tau_2) \quad (3)$$

$$\bar{K}_{i,j}(\tau_1, \tau_2) = \begin{cases} 0 & \text{if } i \text{ and } j \text{ are in the same order in } \tau_1 \text{ and } \tau_2 \\ 1 & \text{if } i \text{ and } j \text{ are in the reverse order in } \tau_1 \text{ and } \tau_2 \end{cases} \quad (4)$$

The Kendall tau rank distance was also expanded to a weighted version called KTDispSq [41]. The same study compared four rank quality assessment metrics and found that the Canberra distance was more correlated to a supervised metric when compared to the original Kendall tau. However, in another analysis, the authors found that Kendall tau outperformed the Canberra distance, remaining unclear which one is the most appropriate metric for their considered problem domain.

### 3.3. Related aggregation problems

Aggregating different information to understand and prioritize decisions, just like the case of risk mitigation of vulnerabilities, is a strategy employed in distinct problems and applications. For instance, rank aggregation methods (Section 3.1) have effectively combined information from different Internet search engines or other databases [8].

In the biological field, data comes from different technological platforms measuring different biological system attributes. Thus, rank aggregation methods have proved to be helpful in this application [8]. Studying different data lists can integrate them and achieve more reliable results. In the work by Lin [8], ranking aggregation was used to combine ranking lists from individual biological studies into an overall list, avoiding dealing with data from different platforms. Colombelli et al. [42] used ranking aggregation within their ensemble feature selection framework for finding genomic candidate biomarkers. Aggregation can also be used in machine learning interpretability, including the analysis of cancer genomic data [43].

## 4. Multi-objective ranking

Real-world problems are usually complex, leading to multiple, sometimes conflicting objectives that must be optimized simultaneously. In this scenario, it is essential to have a trade-off between them [12]. Most of the research focuses on optimizing a single objective, given that it is a more straightforward process. However, this may require artificially engineering a single global objective to represent the many true objectives of the problem [12], which can create setbacks.

### 4.1. Multi-objective optimization

An alternative can be found in multi-objective optimization, in which the multiple objectives are optimized simultaneously. Usually, in such problems, there is no single solution capable of optimizing all objectives. This means that there are several *optimal* solutions because no other solution is better than them considering all the objectives. For these solutions, it is impossible to improve one of the objectives without worsening one or more objectives. This set of optimal solutions is called non-dominated or Pareto optimal solutions, illustrated in Figure 3. Without external criteria (for instance, the user preference), all the non-dominated solutions are considered equally good.

Examples of real-world computing applications of multi-objective optimization are the release plan rescheduling for agile software development [44], efficient traffic routing [45, 46], and topic modeling in text analysis to extract underlying topics from document collections [47]. In practice, it is hard to scale the current algorithms for problems with more than three objectives (called many-objective problems) [12]. Popular multi-objective algorithms are the Non-dominated Sorting GA-II (NSGA-II) by Deb et al. [48] and the Optimistic Linear Support (OLS) by Roijers et al. [49].

It is often said that problems do not need to be modeled with a multi-objective perspective [12]. From this view, it should be possible for all utility functions to be represented by a single scalar signal [12]. The utility function is a function that measures the user's satisfaction regarding the outcome of the system [11]. In the case of risk prioritization, this implies that all risk metrics could be combined into a single global metric. Such an idea could be used to rank vulnerabilities. This option would be somewhat similar to the idea of rank aggregation presented in Section 3.1, but as discussed, those techniques present some limitations. Moreover, creating a single metric by an *a priori* scalarization function is sometimes undesirable, infeasible, or even impossible. It may oversimplify the underlying problem and produce suboptimal results [12].

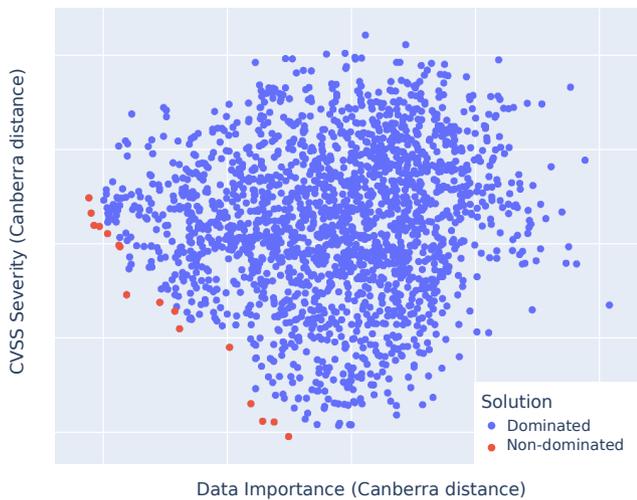

**Figure 3: Example of a Pareto front.** An illustration of the possible solutions in the fitness space for two objectives. The solutions are vulnerability ranks minimizing the Canberra distance (Section 3.2) from a perfect ordered rank according to their CVSS Severity (y-axis) and from a perfect ordered rank according to the Data Importance (x-axis) of data stored in the servers where the vulnerability was found. Solutions in red are non-dominated, representing the Pareto front, while solutions in blue are dominated. The solutions in this example were optimized with the NSGA-II algorithm using authors' private data sources.

Such an idea is applied in the framework proposed by Farris et al. [14], presented in Section 2.2, in which goal programming [50] is used to select the set of vulnerabilities to be mitigated in a period. A goal programming implementation with *a priori* scalarization is used, which sets target values for each objective and weights for the deviation values of each target [14]. These multiple goals are transformed into constraints, and the deviation values act as relaxation variables for the objective target values. The final objective function becomes the minimization of the weighted sum of the deviation values, thus delivering a single definitive solution instead of a set of non-dominated ones. This is a valuable technique for decision support scenarios, where a single final policy must be implemented. Nonetheless, as discussed above, this approach may oversimplify the considered problem domain, requiring user-defined preferences prior to the optimization phase, which must be executed whenever the preferences change.

Another work about risk prioritization using multi-objectives was written by Viduto et al. [15] and presented in Section 2.2. Their approach was to use Multi-objective Tabu Search (MOTS) to minimize the total investment cost and risk of a vector of vulnerabilities. MOTS starts with a random solution and iterates over the solution's neighborhood using the concept of a tabu list [51]. A new solution is selected from the neighborhood if it has the lowest cost or risk within this set. The Pareto front is built by recording all solutions visited and removing the dominated ones. According to Viduto et al. [15], MOTS is faster than other meta-heuristics and demonstrates a good approximation of optimal solutions.

### 4.2. Preference-based multi-objective optimization

According to Roijers and Whiteson [11], there are three scenarios in which it is not advisable to create *a priori* scalarization functions to make the conversion from multi to single objective. When these preferences are available, it becomes possible to select the best option without rerunning the expensive optimization step. The scenarios mentioned previously are the unknown utility function scenario (a), the decision support scenario (b), and the known utility function scenario (c). Additionally, Hayes et al. [12] proposed three more scenarios, the interactive decision support scenario (d), the dynamic utility function scenario (e), and the review and adjust scenario (f). A complete description of each one of these scenarios can be found in the work by Hayes et al. [12].

We argue that, out of the six scenarios presented by Hayes et al. [12], risk prioritization should fit either in the unknown utility function scenario (a) or the decision support scenario (b). These scenarios are illustrated in Figure 4. In both, using *a priori* scalarization is undesirable, which matches the expectation that the risk metrics should not simply be mixed to obtain a global rank of vulnerabilities risk.

In the first scenario, the unknown utility function scenario (a) [52], *a priori* scalarization is unattainable because the utility function is unknown during optimization. There is uncertainty around the utility that could be received from the optimization, and it is better to compute a coverage set of solutions that allows for quick updates in the utility when the context changes [11].

When prioritizing which vulnerabilities should be patched, the severity of the vulnerability and the importance of the server can often be at odds. This means that the objectives can be conflicting. Specifying the exact preferences for these objectives is difficult since certain circumstances can change their priorities. For instance, it may be more important to patch the vulnerabilities with a higher risk in normal circumstances. However, if an attack is detected, the system should first prioritize patching the servers with sensitive data to protect them. In cases like these, the link between these effects and the measures to be taken is not well understood, so learning a set of optimal solutions is essential [11].

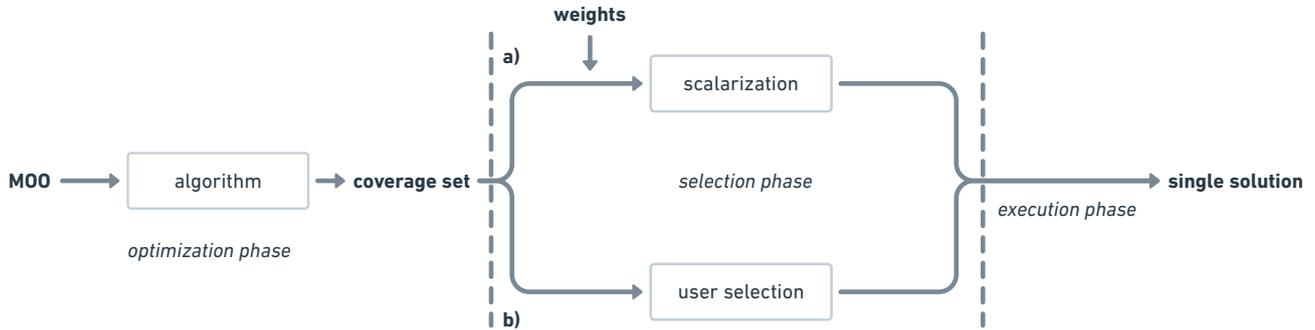

**Figure 4:** The two motivating scenarios for multi-objective optimization (MOO) relevant for the risk prioritization problem: (a) the unknown utility function scenario, (b) the decision support scenario. The main difference between them happens in the "selection phase", in which a solution from the coverage set obtained from the optimization must be chosen. In the first scenario, a solution is found by scalarizing the options using a set of weights for each objective, signaling their importance. In the second scenario, the coverage set is presented to the users, who must decide which solution better suits their needs.

The second scenario relevant to the problem of risk prioritization is the decision support scenario (b). In this case, it is difficult to specify the user's preferences, or they are unknown, thus rendering *a priori* scalarization unfeasible. This scenario is very similar to the unknown utility function one. However, in the selection phase, a set of solutions is presented to the users to select preferences [11].

There is no single optimal solution to specify the risks in risk prioritization, as each stakeholder may have distinct interests or objectives. Thus, the solution depends on each stakeholder's risk management preferences. To accurately model each user's preferences and consider the trade-offs of the distinct objectives would be impractical or impossible. A better option would be to learn a set of optimal solutions and then decide which one should be used following the collective decision of the users [11].

Although the two scenarios presented above are similar, they have a main difference regarding the selection phase. There is a revelation step in which the utility function is made explicit in the unknown utility function scenario [12]. Meanwhile, the users decide in the decision support scenario, while the utility function is implicit in the decision. It is hard to define a utility function explicitly, so the decision support scenario is often the chosen option for implementation [12].

### 4.3. Improvements and applications

Because the number of Pareto non-dominated solutions grows as more objectives are added to the problem, some research efforts are proposing methods for ranking the non-dominated solutions, as reviewed by Garza-Fabre et al. [53]. This is relevant because, in a scenario where there are multiple equally significant objectives, for example, 20, it is reasonable to say that a solution $X_i$ better than $X_j$ in 19 objectives and worse in only one, is also overall a better solution, despite both being possibly included in the Pareto front. Nonetheless, in some scenarios, such as the unknown utility function, there will be a revelation step where weights can be assigned to a utility function after the optimization phase. Depending on such weights, $X_i$ could be a worse solution than $X_j$, even though $X_i$ better fits most objectives.

Instead of ranking the various non-dominated solutions in a many-objective optimization problem, an approach employed by some reinforcement learning algorithms is the Thresholded Lexicographical Ordering (TLO) [54]. It aims to rescue the problem's feasibility by eliminating extreme solutions that do not provide a minimal fitness for all the objectives. These thresholds can be set manually by an extensive analysis or expert opinions [55] and deduced automatically through the learning phase [56]. On the other hand, the lexicographical ordering can be used to select a final single best solution among those that survived the imposed thresholds for each objective.

While the TLO technique can be employed to mine the Pareto front searching for a small set or a specific solution, which is desirable in decision support scenarios, the thresholds can also restrict the power of freely indicating the importance of each objective. Consider, for example, a particular situation after the optimization phase in which just one objective is relevant, and the others become entirely irrelevant. In this case, the thresholds could have eliminated the best possible solution for the current user needs due to the unreached minimal values for the other objectives that turned out to be irrelevant.

Considering what was presented above, it would be advantageous to model the problem of risk prioritization as a multi-objective task, in which there are a few metrics regarding the risk that will be used to create the objectives. Unlike *a priori* scalarization or rank aggregation, this approach would result in a coverage set of optimal ranks of the risk without losing the multi-factors that determine the severity and urgency of a vulnerability. The users would have more control over the utility function, allowing for quick updates in the strategy and an overall more dynamic system, able to adapt to changes in the environment, being they sudden (for instance, reacting to an attack) or slow (a shift in the management strategies).

Although there is not much research in using multi-objective for the risk prioritization of vulnerabilities, there are works that apply multi-objective optimization for ranks. Carrillo and Jorge [57] and Momma et al. [58] proposed using common-weight Data Envelopment Analysis and Augmented Lagrangian based method to build ranks, respectively. Streimikiene and Balezentis [59] used the MULTI-MOORA algorithm to prioritize and rank climate change mitigation strategies. Finally, Dalal et al. [60] employed Hodge decomposition to rank comments on the web.

### 4.4. Model suggestion

To successfully apply the MOO approach, it is imperative to have a good problem representation. In this subsection, a template on how to implement the MOO for vulnerability prioritization is presented. This should serve as a guideline and initial step for actual implementation and does not consider the several specificities of each security scenario.

The fitness functions for each objective should be computationally efficient and expressive enough to assess each feasible solution's quality properly. For instance, an inversion count yields a high computational load to the optimization, so using the rank distance metrics from Section 3.2 would be more efficient. The solution array can be represented as a rank-based vector $R$ in which the element $i$ is the ranking of the $i$th vulnerability in the list of vulnerabilities. Thus, there are as many decision variables as vulnerabilities in the dataset, and their ranking values are lower bounded by 1 (first position in the priority rank) and upper bounded by the number of vulnerabilities $n$. For example, the solution vector $[4, 1, 3, 2]$ assigns the priority ranking of 4 to the dataset's first vulnerability, 1 to the second, 3 to the third, and 2 to the fourth. In population-based algorithms such as the NSGA-II, each individual would be one of these rank-based vectors, and the mutation and crossover operations would modify the ranking of the vulnerabilities inside this vector.

For a particular metric $m$, the perfect rank sorts all vulnerabilities according to their $m$ values. Because a company may be interested in prioritizing the vulnerabilities according to multiple metrics, each metric $m$ will represent an objective $O_m$ and have its perfect rank $P_m$. Note that these perfect ranks are feasible solutions that lie in the extremities of the Pareto front (because they optimize one of the objectives irrespective of the others).

The optimization goal is to minimize the distance between a possible solution $R$ and each $P_m$ using a particular distance function (Section 3.2). Because the notation of rank distance and minimization are being used, the Kendall tau and Spearman's rank correlation output needs to be multiplied by −1. This is due to these functions outputting 1 if two ranks are identical. Therefore, assuming the use of the Spearman's rank correlation as the function $r_s(p, q)$ (Equation 2) for measuring the distance between the ranks $R$ and $P_m$, the fitness $f_m^R$ corresponding to the objective $O_m$ of an individual $R$ is computed by the following equation:

$$f_m^R = -r_s(R, P_m) \qquad (5)$$

Some open questions with several possible improvements are: how to represent the list of vulnerabilities and model the optimization problem. Other fitness functions and optimization algorithms can be explored. Newer algorithms, such as the Adaptive Geometry Estimation based Multi-Objective Evolutionary Algorithm (AGE-MOEA) [61], could better explore the search space since it estimates the geometry of the Pareto front and adapts the diversity and proximity metrics. The ranks' representation can be modified to be more efficient, for instance, removing vulnerabilities with the same values (as there cannot be a clear preference among them), and the use of initialization and post-optimization strategies could improve the results.

## 5. Research opportunities and trends

As seen in Section 2.2, there are several proposed metrics or strategies on how to compute the risk score of a vulnerability. In addition to these custom metrics, there are also metrics closely tied to risk assessment environments. For instance, one could consider the importance of the data stored in a server, the connectivity of a specific server in the network, or the time passed since the vulnerability discovery as relevant to the overall vulnerability risk.

According to Jacobs et al. [16], the security risk cannot be reduced to a single value like the CVSS score, nor are such scores capable of containing the entirety of the security risk. If there were two metrics available, one for risk severity (the CVSS score) and the other measuring the probability of a risk being exploited, one option would be to scalarize these two metrics in a new single metric, for instance, by multiplying them. This approach could look like a better measurement of risk, but just applying mathematical operations to combine them is faulty and should be avoided [16].

In this context, in which several metrics offer orthogonal information about the risk, it would be ideal to consider their values simultaneously but individually [16]. As seen in Sections 3 and 4, alternatives can include ranking the risk of each vulnerability by each metric and aggregating them or creating a new metric that encompasses the others. However, there are several issues associated with these strategies. First, it prevents the stakeholders, who should make the decisions, from evaluating well-informed trade-offs. Second, the decision-making process is less interpretable and less transparent. The process becomes semi-manual. It does not account for the different preferences distinct users may have, nor that these preferences may change over time [12]. Moreover, these strategies may lead to a loss of information needed to understand or evaluate the solutions [12].

Due to these reasons, viewing the problem of risk prioritization from a multi-objective perspective may bring several advantages; however, this topic has been neglected in the literature [17]. The methods and arguments presented in Section 4 should be a starting point on how and why multi-objective can be used to solve this problem. By considering each metric associated with risk as an objective to be optimized, a multi-objective algorithm can return a set of optimal risk prioritization ranks, from which it would be possible for the users to choose a solution that satisfies their current needs. This choice could account for distinct needs for each user and, as the context changes, be dynamically and quickly updated.

In Section 4, a framework for vulnerability prioritization proposed by Farris et al. [14] is discussed from a multi-objective optimization perspective. The system successfully integrated output data from a vulnerability scanner tool with user-defined preferences to consider multiple possible prioritization scenarios, recommending the best set of vulnerabilities to be patched for that particular context in a period. However, the algorithm relies on a set of *a priori* preferences chosen before the optimization phase begins, including target values for each objective, weights for each deviation to be minimized, the estimated number of personnel-hours required to mitigate a vulnerability, among others. The optimization aims to generate a final solution with a minimum weighted sum of the resulted deviation values, which is presented as a set of vulnerabilities to be mitigated.

Nonetheless, the set of selected vulnerabilities, and the other vulnerabilities present in the scanned system, are not ranked, representing limitations such as which vulnerabilities, among the selected ones, are to be mitigated first. Even though their mitigation utility rank could be used, the rank itself is not optimized. Instead, it is built based on a weighted sum of the considered risk attributes, where the weights are also user-defined. Additionally, if the user preferences change or all the selected vulnerabilities are mitigated before the given time window expires, a new framework execution must be performed in the target system to select new vulnerabilities. Evaluating this approach, Shah et al. [62] compared individual attribute value optimization and multiple attribute value optimization policies and concluded that each has its advantages depending on the scenario and that multiple attributes can achieve balanced performances while mitigating vulnerabilities.

Another multi-objective problem was formulated by Viduto et al. [15] to review the solutions with a good balance between risk and cost in network security. Based on their results, a good perspective for the multi-objective approach was introduced for unbiased decision-making in the vulnerabilities scenario where more than one objective needs to be balanced. According to the authors, this approach allowed the model to accomplish more justified and informed decisions.

As a result of this discussion, several challenges can be perceived. Le et al. [2] observed that the vulnerability assessment field has a significant improvement possibility, mainly regarding the availability of data sources and data-driven models. Nevertheless, understanding how to apply software vulnerability assessment and prioritization can become a challenge if real-world scenarios (academia and production) are not considered. Farris et al. [14] acknowledged that organizational metrics employed in vulnerability evaluations should be previously studied to avoid unintended adverse effects on the analysis results. Consequently, more methods implementing structured vulnerability response programs and solutions to better manage the false-positive rates in vulnerabilities scanners outputs would be welcome to achieve more accurate analysis.

Hayes et al. [12] and Saini and Saha [13] presented one final provocation for future works that refers to the limitation of the current algorithms regarding the number of objectives (many-objectives problems). Usually, the problems become infeasible for a number much higher than three [53]. One alternative that could be explored in this scenario is the use of hybrid methods, in which similar metrics are combined, still maintaining more than one metric to be used as objectives. An arising challenge would refer to combining metrics without violating optimality guarantees. In that case, theoretical investigations on performance guarantees would represent another promising direction.

## 6. Conclusion

The main goal of this review was to draw attention to the problem of risk prioritization for vulnerabilities. This is a relevant security problem with several approaches in academia and industry. Managing vulnerabilities is an essential task; however, a dependency on resources creates the need for a strategy that enables dealing with numerous vulnerabilities. Vulnerabilities with the highest risk should be dealt first and, if the resources limit is reached, the remaining vulnerabilities should be postponed for the next iteration. A risk prioritization rank would be the best solution to ensure that the most critical vulnerabilities are managed first.

As seen, most of the available works focus on modeling the prioritization from the more common single-objective perspective. Based on the analyzed work, the drawbacks, and limitations of this kind of strategy were presented. The main contributions of this review consist of an example of a vulnerability management life cycle, a summary of the most reliable techniques that can be employed to elaborate the risk prioritization rank, a discussion about overall challenges to compute the risk score of a vulnerability and the feasibility of using a multi-objective proposition to prioritize vulnerabilities. These contributions serve as a guideline for other academic research in this area and offer optimized strategies that entities can consider when handling their vulnerability management life cycles, ensuring that the vulnerabilities with higher risks will always be dealt with first, so in a postponement scenario, only the less critical vulnerabilities will be postponed. This will also affect the final users; once the most critical vulnerabilities receive patches, they will be less exposed to the most dangerous threats.

This work proposes using multi-objective optimization to manage the vulnerabilities risk prioritization, thus creating priority ranks sets by simultaneously considering several relevant metrics. Such an approach represents a novel, relevant research direction that would bring some key advantages to the users, significantly more dynamism regarding changes in their needs or environment, reduced associated costs, optimized time management for companies, and more safety for final users thus should be the focus of further research. This review can be used as a guideline to validate and assess the presented algorithms implementations, metrics, and frameworks with different objectives so that an optimized solution can achieve state-of-the-art vulnerabilities risk ranking.

## CRediT authorship contribution statement

**Bruno Iochins Grisci:** Conceptualization, Methodology, Validation, Formal analysis, Investigation, Writing - Original Draft, Visualization. **Gabriela Kuhn:** Conceptualization, Methodology, Validation, Formal analysis, Investigation, Writing - Original Draft. **Felipe Colombelli:** Conceptualization, Methodology, Validation, Formal analysis, Investigation, Writing - Original Draft, Visualization. **Vítor Kehl Matter:** Conceptualization, Methodology, Validation, Formal analysis, Investigation, Writing - Original Draft, Visualization. **Leomar Lima:** Conceptualization, Data Curation. **Karine Heinen:** Conceptualization. **Mauricio Pegoraro:** Conceptualization. **Marcio Borges:** Conceptualization. **Sandro José Rigo:** Validation, Formal analysis, Writing - Review & Editing. **Jorge Luis Victória Barbosa:** Validation, Formal analysis, Writing - Review & Editing. **Rodrigo da Rosa Righi:** Validation, Formal analysis, Writing - Review & Editing. **Cristiano André da Costa:** Validation, Formal analysis, Writing - Review & Editing. **Gabriel de Oliveira Ramos:** Conceptualization, Methodology, Validation, Formal analysis, Writing - Review & Editing, Supervision, Project administration, Funding acquisition.

## Declaration of competing interest

The authors declare that they have no known competing financial interests or personal relationships that could have appeared to influence the work reported in this paper.

## Acknowledgements

This study was financed in part by the Coordenação de Aperfeiçoamento de Pessoal de Nível Superior - Brasil (CAPES) - Finance Code 001, Conselho Nacional de Desenvolvimento Científico e Tecnológico - CNPq, and Fundação de Amparo à Pesquisa do Estado do Rio Grande do Sul - FAPERGS. This work was supported by Dell Inc. via the 18th Amendment to the Technical and Scientific Cooperation Agreement No. 01/2017 – Information Technology Innovation Support Law – Brazilian Government. The authors also would like to thank Universidade do Vale do Rio dos Sinos (Unisinos) and the Graduate Program in Applied Computing (PPGCA) for providing the support and infrastructure required for the project development.